\documentclass[12pt]{article}

\newcommand{\sect}[1]{ \section{#1} } 
\newcommand{\subs}[1]{\subsection{#1}}

\newcommand{\ve}{\left( \begin{array}{r}} 
\newcommand{\ev}{\end{array} \right)} 
\newcommand{\ar}{\left( \begin{array}{rr}} 
\newcommand{\ra}{\end{array} \right)} 
\newcommand{\arr}{\left( \begin{array}{rrrr}} 
\newcommand{\arrr}{\left( \begin{array}{rrrrrr}} 

\newcommand{\eqr}{\begin{eqnarray}}
\newcommand{\rqe}{\end{eqnarray}} 
\newcommand{\eq}{\begin{equation}} 
\newcommand{\qe}{\end{equation}}

\newcommand{\half}{\frac{1}{2}}

\newcommand{\adss}{\mbox{$AdS_5\times S^5$}}
\newcommand{\ads}{\mbox{$AdS_5$}}
\newcommand{\rep}[3]{({\bf #1}, {\bf #2})_{#3}}

\def\KK{{\rm I\kern -.2em  K}}
\def\NN{{\rm I\kern -.16em N}}
\def\RR{{\rm I\kern -.2em  R}}
\def\ZZ{Z}
\def\ZZZ{{\small{\rm Z}\kern -.5em Z}}
\def\QQ{{\rm \kern .25em
             \vrule height1.4ex depth-.12ex width.06em\kern-.31em Q}}
\def\CC{{\rm \kern .25em
             \vrule height1.4ex depth-.12ex width.06em\kern-.31em C}}

\title{Large $N$ Superconformal Gauge Theories and Supergravity
Orientifolds} 
\author{Ansar Fayyazuddin\footnote{email: ansar@schwinger.harvard.edu} 
\quad and \quad
Micha\l\ Spali\'nski\footnote{email: mspal@schwinger.harvard.edu} 
\footnote{On leave from the Institute of Theoretical Physics, Warsaw
University.}}
\date{}

\begin{document}

\maketitle

\begin{center}

{\em 
Lyman Laboratory of Physics\\
Harvard University\\
Cambridge, MA 02138, U.S.A. 
}

\end{center}

\vspace{1.4cm}

\begin{abstract} 

Maldacena's duality between conformal field theories and supergravity is
applied to some conformal invariant models with $8$ supercharges appearing
in the F-theory moduli space on a locus of constant coupling. This includes
$Sp(2N)$ gauge theories describing the worldvolume dynamics of D3-branes in
the presence of D7-branes and an orientifold plane. Other examples of this
kind are models with exceptional global symmetries which have no
perturbative field theory description. In all these cases the duality is
used to describe perturbations by primary marginal and relevant operators.

\end{abstract} 

\vspace{-16cm}
\begin{flushright}
HUTP-98/A037\\
hep-th/9805096\\
\end{flushright}

\thispagestyle{empty}

\newpage

\setcounter{page}{1}

\sect{Introduction}

Maldacena has proposed a new kind of ``duality'' between large N conformal
field  
theories in various dimensions and supergravity on anti-de Sitter spaces
\cite{mal}.  
The idea advocated in \cite{mal} is based on the construction of 
gauge theory as the world volume theory of N D-branes in the $\alpha
'\rightarrow 0$ limit. The limit is taken in such a way as to keep the
masses of field theory excitations finite.  On the other hand this system
can be described by the black brane solutions of supergravity.  In the
$\alpha '\rightarrow 0$ limit one is probing the near-horizon geometry of
the black branes which is typically an anti-de Sitter space times a compact
manifold \cite{gib82,kalp,fergk}.  Since supergravity is expected to
provide a reliable description 
only if the curvature is small (so $\alpha^\prime$ corrections can be
neglected), one is lead to consider a limit in which the
radius of curvature is large which corresponds to the limit of a large
number of branes. 
This observation has attracted a great deal of attention. Apart from
comparing the spectra, it was also possible to calculate correlation
functions in the conformal field theory using this
correspondence\cite{with,gubkp,fri}.

The original conjecture referred to theories with 16 supersymmetries,
however it was quickly realized that one could also use it in theories with
less supersymmetry \cite{kacs,lawnv,berkv,ozt,berj,f1,f2}. 
Here a family of superconformal field theories with $8$ supersymmetries
is considered: the $Sp(2N)$ gauge theory with an antisymmetric tensor
hypermultiplet and four fundamental hypermultiplets. 
This system is an orientifold of type IIB theory in which D7-branes are
localized at the orientifold fixed planes with D3-branes. The field theory
of interest describes the D3-brane dynamics. 
This model has a vanishing beta function and is believed to be
exactly superconformal. Since
the field theory description is known, one can analyze the spectrum and
make a comparison much as in \cite{with,ozt}. The theory has a global
$SO(8)$ symmetry, due to gauge symmetry enhancement in the D7-brane
worldvolume. 

This theory can also be regarded as a particular F-theory
compactification in a region where the coupling is
constant\cite{sen,dasm,sei,bands,douls,leey}. This way of looking at it is
revealing, and 
it leads one to consider other superconformal quantum systems, which do not 
have any known field theory description. While in the $Sp(2N)$ case the
orientifold involves a $\ZZ_2$ reflection, the cases with $\ZZ_3$, $\ZZ_4$, $\ZZ_6$
lead to worldvolume field theories on the D3-branes with E$_6$, E$_7$,
E$_8$ global symmetry\cite{witcom,minn1,minn2}. While these theories have no
known field theory 
description, they can in the present context be treated in the same way as
the $\ZZ_2$ case which does.  
All of these theories can be obtained by
compactifying an M-theory 5-brane \cite{ganms} on a torus  
in the presence of an end-of-the-world 9-brane.  The 5-brane theory has 
$E_8$ global symmetry due to the presence of the $E_8$ gauge symmetry
living on the 9-brane.  This suggests the possibility that the 
non-perturbative 4-dimensional $E_n$ theories obtained from 
the six-dimensional theory may not be ordinary field theories but could
inherit tension-less strings from its progenitor.   

In the limit where the correspondence advocated in \cite{mal,gubkp,with} is
expected to hold these theories turn out to be described by $\ZZ_n$
orientifolds\footnote{Orbifolds of \adss\
have recently been discussed in \cite{kacs,lawnv,berkv,ozt,berj,f1,f2}, and
orientifolds in  
\cite{kak1,kak2}.} of \adss. The twist acts only on the $S^5$ factor, so
that the 
resulting theory on the boundary of \ads\ is conformal. It is
straightforward to analyze what happens to the Kaluza-Klein spectrum of the
${\cal N}=4$ theory on \adss\ and to identify the chiral primary
operators. This 
note presents the results for the low-lying scalar operators. In the
$Sp(2N)$ case it is also easy to identify the chiral primaries in terms of
the fields appearing in the perturbative description, using their 
group theoretical properties which follow from the calculation.

While the $Sp(2N)$ case is fairly well understood since it has 
a perturbative formulation\cite{sen}, the remaining cases have no
known field theory 
description: they are inherently non-perturbative.  Nevertheless, the
methods developed in \cite{with,ozt} allow one to identify the quantum
numbers of the relevant and marginal perturbations of these theories even
though no gauge field theory description is known. 

Due to multiplet shortening \cite{flaf,fro,angffs} it is expected that
information about the spectrum of primary operators should be exact. In
this sense supergravity allows the study of inherently non-perturbative
phenomena on the gauge theory side which are inaccessible using standard
field theory methods.

\sect{F-theory Orientifolds with Constant Coupling}

F-theory \cite{vafa} is believed to describe non-perturbative 
type IIB theory.  It is a powerful method for generating non-trivial
type IIB backgrounds with varying dilaton and Ramond-Ramond scalar fields.
By definition F-theory on an elliptically fibered manifold is type
IIB string theory on the base manifold of the fibration with the 
complexified 
coupling $\tau = \chi + i\exp{(-\phi)}$ identified with the modular
parameter of the elliptic fiber.  At a point on the base where
the fiber degenerates one has a localized 7-brane. 

In \cite{sen,dasm} F-theory compactifications on elliptic K3 with
constant type IIB coupling were considered.  These K3's are described by
orbifolds $T^4/\ZZ_n$ with $n=2,3,4,6$ and give 8-dimensional gauge symmetry
of SO(8), $E_6$, $E_7$, $E_8$ respectively which can be understood from the
point of view of 
coincident 7-branes at the orientifold point \cite{pol,gz,ghz,im}.  
The base of the elliptic fibration is an orbifold $T^2/\ZZ_n$. The fiber
above a fixed point degenerates and there is a nontrivial monodromy in the
fiber\cite{sen,dasm}, which means that apart from the $\ZZ_n$ projection on
the base there 
is also a $\ZZ_n$ action on the complexified coupling (dilaton and axion
pair) and antisymmetric field (NS-NS and R-R two-form gauge fields). This
monodromy is $-1$ in the $\ZZ_2$ case, and $S$, $ST$ and $(ST)^2$ in the
$\ZZ_3$,  $\ZZ_4$ and $\ZZ_6$ cases. 

These models can also be described as supergravity solutions with metric
\cite{gsvy}: 
\eq
ds^{2}=\tau_{2}\mid \eta^{2}({\tau})\Delta^{-1/12}dz\mid^{2} +\delta_{ij}
dx_{i}dx_{j},
\qe
where $\tau_2$ is the imaginary part of the coupling $\tau$, $\Delta$ is
the discriminant of the elliptic fiber
\eq
\Delta(z)=4 f^3(z) + 27 g^2(z)
\qe
and $\eta$ is the Dedekind
function.  The metric is modular invariant.  The K3s are described by the
Weirstrass equation for the elliptic fiber as a function of $z$, the
coordinate on the base manifold $P^1$:
\eq
y^2 = x^3 + f(z) x + g(z)
\qe
with $f$ a polynomial of degree $8$ and g a polynomial of degree $12$
\cite{vafa}.
$\tau$ is determined from the $j$-invariant of the equation:
\eqr
j(\tau(z)) & = &\frac{4 (24 f(z))^3}{\Delta (z)} \nonumber \\
& = &
\frac{(\theta^{8}_{1}(\tau)+\theta^{8}_{2}(\tau)+\theta^{8}_{3}(\tau))^3}
{\eta^{24}(\tau)}.
\rqe
K3 manifolds with constant $\tau$ are the ones for which $j$ is constant. 
There are 4 cases for which a type IIB orientifold description
exists \cite{sen,dasm}:

\vskip 0.3cm

(i) $T^4/\ZZ_2$:
\eqr
f(z) & = & \alpha \prod_{i=1}^{4}(z-z_i)^2 ,\nonumber \\
g(z) & = & \prod_{i=1}^{4}(z-z_i)^3 .
\rqe
The coupling constant depends on the parameter $\alpha$ and can be fixed
to be any value by choosing an appropriate $\alpha$.
In this case there are $4$ points ($z=z_i$) at which the fiber
degenerates. 
The K3 has four $D_4$ singularities resulting in the gauge group
$SO(8)^4$.  Each $SO(8)$ can be understood from the type IIB point of view
as 4 D7-branes 
coinciding with an orientifold fixed plane \cite{sen}.

\vskip 0.3cm

(ii) $T^4/\ZZ_3$
\eqr
f(z) & = & 0 , \nonumber \\
g(z) & = & \prod_{i=1}^{3}(z-z_i)^4 .
\rqe
In this case\cite{dasm}, $j=0$ thus the coupling constant is fixed to be 
$\tau = \exp i\pi /3$.  The singularity type close to $z=z_i$ is
$E_6$, resulting in the gauge group $E_6\times E_6\times E_6$.  

\vskip 0.3cm

(iii) $T^4/\ZZ_4$
\eqr
f(z) & = & (z-z_1)^3(z-z_2)^3(z-z_3)^2 ,\nonumber \\
g(z) & = & 0 .
\rqe
In this case \cite{dasm}, $j=(24)^3$; thus the coupling constant is fixed
to be  
$\tau = i$.  The singularity type close to $z=z_1,z_2$ is
$E_7$ and close to $z=z_3$ it is $D_4$ resulting in the gauge group 
$E_7\times E_7\times SO(8)$.  

\vskip 0.3cm

(iv) $T^4/\ZZ_6$
\eqr
f(z) & = & 0 , \nonumber \\
g(z) & = & (z-z_{1})^{5}(z-z_{2})^{4}(z-z_{3})^{2} .
\rqe
In this case \cite{dasm}, $j= 0$; thus the coupling constant is fixed to be  
$\tau = \exp i\pi /3$.  The singularity type close to $z=z_1$ is
$E_8$, close to $z=z_2$ it is $E_6$ and close to $z=z_3$ 
it is $D_4$, resulting in the gauge group $E_8\times E_6\times SO(8)$.  

The gauge groups above can also be understood using Chan-Paton factor
analysis of coinciding 7-branes using multi-pronged strings
\cite{gz,ghz,im}.

Let us focus our attention on the vicinity of one of the points where the
elliptic fiber degenerates, i.e. where the 7-branes are localized.  One can
introduce D3-branes \cite{bands} in the above backgrounds and reduce the
supersymmetry to $8$ real supersymmetries corresponding to ${\cal N}=2$ in
$4$ dimensions.  The theory living on the 3-branes will be an ${\cal N}=2$
Seiberg-Witten theory with the same flavour symmetry group as the 7-brane
gauge group \cite{bands}.  As discussed above the Seiberg-Witten theory
of the 3-brane probing the 7-brane geometry can have a perturbative 
description only in case (i).  

If one introduces $N$ 3-branes close to a $D_4$ singularity and takes the
other singularities to be far away, then the resulting theory on the
probe is a ${\cal N}=2$ gauge theory with gauge group $Sp(2N)$ coupled to a
second rank anti-symmetric tensor hypermultiplet and $4$ fundamental
hypermultiplets \cite{douls}.  The four fundamental hypermultiplets arise
from strings stretched between the D7-branes and the D3-branes.  The
anti-symmetric tensor arises from the action of the orientifold projection
on the fields transverse to the D3-branes but parallel to the D7-branes.
The theory has a Coulomb branch parameterized by the expectation value of
the complex scalar field in the ${\cal N}=2$ vector multiplet and
corresponds to moving D3-branes away from the fixed plane.
There is also a Higgs branch parameterized by the expectation values of the
two complex scalars in the anti-symmetric tensor hypermultiplet, they
correspond to the motion of a subset of the D3-branes parallel to the
D7-branes.  Finally, there is a second Higgs branch which is parameterized 
by the scalars in the 
fundamental representation and corresponds to dissolving some D3-branes
inside D7-branes.

\sect{Supergravity Solutions}

This section describes adding D3-branes to the D7-brane backgrounds
described above. Let us take N D3-branes with the worldvolume along $x^0,
x^1, x^2, x^3$ and the appropriate number of D7-branes (according to the
discussion in the previous section) with worldvolumes along $x^0, x^1,
\dots, x^7$. Let $z\equiv x^8+ix^9, v\equiv x^4+ix^5, \tilde{v}\equiv
x^6+ix^7$.  The position of the D7-brane is given by $z$, while the
D3-brane position is given by $z,v,\tilde{v}$. From the point of view of
the field theory on the D3-branes $z$ parameterizes the Coulomb branch of
the theory while $v,\tilde{v}$ parameterize the Higgs branch. Thus $z$
corresponds to the vacuum expectation value of the adjoint Higgs, while
$v,\tilde{v}$ correspond to the expectation values of the ${\cal N}=1$
chiral multiplets comprising the hypermultiplet in the antisymmetric
representation of the gauge group.

Let us focus our attention on the vicinity of an 
F-theory K3 singularity which corresponds to coinciding 7-branes in type
IIB theory.  For the present cases, where $\tau$ is a constant, the
part of the metric describing the D7-branes\footnote{In the exceptional
cases these are actually (p,q) 7-branes}  (and $\Omega_7$) backgrounds 
with zero net 7-brane charge is (up to a constant normalization): 
\eqr 
\mbox{SO(8):}\quad ds^2 & = &\mid z^{-\frac{1}{2}}dz\mid^2 ,\nonumber \\ 
\mbox{$E_6$:}\quad ds^2 & = &\mid z^{-\frac{2}{3}}dz\mid^2 ,\nonumber \\ 
\mbox{$E_7$:}\quad ds^2 & = &\mid z^{-\frac{3}{4}}dz\mid^2 ,\nonumber \\ 
\mbox{$E_8$:}\quad ds^2 & = &\mid z^{-\frac{5}{6}}dz\mid^2. 
\rqe 
These can be described as orbifolds $\CC/\ZZ_n$ with $n=2,3,4,6$
respectively.  The covering space is the complex u-plane, with $z =
u^n$. In these coordinates the metric is $ds^2 = \mid du\mid^2$. The action
of the orientifold group on $u$ is $u\longrightarrow \exp{(2\pi i/n)} u$.  

It is easy to introduce 3-branes into the
problem: this is done in such a way as to respect the identification
under the orientifold group. The black-brane solution for 3-branes of
\cite{hors} reads 
\eqr
ds^2 & = &f^{-1/2}dx_{\parallel}^2 + f^{1/2} dx_\bot^2 ,\\
f & = &1 + \frac{4\pi gN\alpha '^2}{r^4}\nonumber
\rqe
where
\eqr
dx_{\parallel}^2 &=& -(dx^0)^2 + \sum_{k=0}^{3} (dx^k)^2 , \nonumber \\
dx_\bot^2 &=& \mid du\mid^2 + \mid dv\mid^2  + \mid d\tilde{v}\mid^2.
\rqe
The solution in the presence of the 7-brane --- orientifold system
is obtained simply by identifying $u\rightarrow u\exp (i2\pi/n)$. One can
now express the metric in terms of the single valued $z=u^n$.  
It is convenient to use the following variables:
let $r$ be the 
radial distance away from the point where the $N$ D3-branes are located (in
the $6$ dimensional transverse space), and let $R$ be the projection onto
the 
the covering u-plane, i.e.  $R=r\cos\phi$ with
$\phi\in[0,\pi/2]$. Then one has $z=R^{n}\exp{i\theta}$ with
$\theta\in[0,2\pi]$. 
This leads to 
\eqr 
ds^2 & = &f^{-1/2}dx_{\parallel}^2 + f^{1/2}(dr^2 +
r^2d\phi^2 + \frac{r^2}{n^2}\cos^{2}\phi d\theta^2 + r^2\sin^{2}\phi
d\Omega_{3}^2) ,\nonumber  \\ 
f & = &1 + \frac{4\pi gN\alpha '^2}{r^4} . 
\rqe
Taking the near horizon limit as in \cite{mal}, keeping $r/\alpha^\prime 
\equiv U$ fixed while taking $\alpha '\rightarrow 0$, one arrives at:
\eqr
ds^2 & = &\alpha^\prime \{\frac{U^2}{\sqrt{4\pi gN}}dx_{\parallel}^2 + 
+ \sqrt{4\pi gN}\frac{dU^2}{U^2} + \nonumber \\ 
&+& \sqrt{4\pi gN} (d\phi^2 + 
\frac{\cos^{2}\phi}{n^2}d\theta^2 + 
\sin^{2}\phi d\Omega_{3}^2)\} .
\rqe
This shows that in the near-horizon limit the metric looks like
$AdS_{5}\times S^{5}/\ZZ_n$ as in  
\cite{kacs,lawnv}, except that now the orbifold group action is different.
Consider $SU(2)\times SU(2)\times U(1)\subset SU(4)\sim SO(6)$ where
$SO(6)$ is the isometry group of $S^5$.  In \cite{kacs,lawnv} the orbifold 
action was taken to act on one of the $SU(2)$ factors, whereas in the
present case the
orbifolding acts only on the $U(1)$ factor.  The orbifold action
does not break the $U(1)$ but identifies $\exp (i\theta)$
with $\exp i(\theta + 2\pi /n)$.  

The $SO(9,1)$ Lorentz group is broken by the D7-brane to $SO(7,1)\times
SO(2)$ and the D3-branes break this further to $SO(3,1) \times SO(4)\times 
SO(2)$. Thus the global symmetry 
group of the theory on the D3-brane is 
$SO(3,1) \times SO(4)\times U(1) \sim SO(3,1)\times SU(2)\times SU(2)\times
U(1)$. This is the Lorentz group in $3+1$ dimensions, an $SU(2)\times U(1)$
R-symmetry group (as expected for ${\cal N}=2$ supersymmetry) and an
$SU(2)$ 
(non-R) global symmetry.

\sect{The Spectrum}

The spectrum of type IIB supergravity on \adss\ was found in
\cite{kim,gun}. It 
falls into representations of the $SU(4)$ R-symmetry.  
The states of spin zero families of zero or negative mass can be
summarized by the following table:

\begin{center}
\vspace{.2cm}
\begin{tabular}{|c|c|c|c|c|}
\hline
$\Delta$     & $m^2$    &  Range      & Dynkin Label & Irreps \\ \hline\hline
k   & $k(k-4)$ & $k\geq 2$    &  $(0, k, 0)$ & {\bf 20'}, {\bf 50}, {\bf
105}, $\dots$ \\ 
\hline
k+3   & $(k-1)(k+3)$ & $k\geq 0$    &  $(0, k, 2)$ & {\bf 10}, {\bf
45}, {\bf 126}, $\dots$ \\ 
\hline 
k+4   & $k(k+4)$ & $k\geq 0$    &  $(0, k, 0)$ & {\bf 1}, {\bf 6}, {\bf
20'}, $\dots$ \\ \hline 
\end{tabular}

\vspace{.2cm}
{\bf Table 1}
\end{center}
\vspace{.1cm}

Here $\Delta$ is the dimension of the operator that the field couples to in
the conformal field theory. It is related to the mass through the formula
\cite{with} 
\eq
m^2 = (\Delta + p)(\Delta + p - d)
\qe
for p-form fields in d dimensions, where the correct solution for $\Delta$
is the larger of the two roots.  

The orientifold construction that one is lead to consider for the D3-D7
system 
involves modding out by $\ZZ_n$ subgroups of the $U(1)$ in the decomposition 
$SU(4) \supset SU(2)\times SU(2)\times U(1)$. In addition, the orientifold
acts non-trivially on the supergravity fields. As discussed in section $2$,
it affects the complexified coupling and 
two form gauge fields, 
leaving the other fields invariant\cite{sen,dasm}. This will be
discussed on a case by case basis below. Since it is only the second rank
antisymmetric tensor fields that are non-trivially transformed, this will
only affect the modes coming from 
the representations ${\bf 10}$ and ${\bf 45}$ of $SU(4)$ \cite{kim,gun}.

The $SU(4)$ representations appearing in table $1$ have to be decomposed
under $SU(2)\times SU(2)_R\times U(1)_R$. The action of the orientifold
group is a product of the (spacetime) orbifold phase determined by the
$U(1)_R$ charge and the orientifold action on the supergravity fields. In
the cases where the latter action is trivial the projection leaves the
operators with $Q=0$ mod $2n$ (for $\Gamma=\ZZ_n$). These are all the cases
which do not arise from the two-forms. The operators coming from the
two-forms are in the ${\bf 10}$ and ${\bf 45}$ of $SU(4)$, and there one
has to account for the extra phase from the orientifold group. It turns
out, as discussed below, that the operators which remain after the
projection have $Q_R=\pm2$ mod $2n$. 

This leads to the spectrum of invariant operators in the ${\cal N}=2$
theory. Only some of the resulting operators will be chiral
primaries of the conformal field theory. This is due to the fact, that for
a chiral primary operator\footnote{There will of course be antichiral
primaries also, which are not shown in the tables below.} of dimension
$\Delta$ the following relation must  
be satisfied: 
\eq
\label{charge}
\Delta = \half (R + 2 J) ,
\qe
where $R$ is the R-charge of the operator, and $J$ is its $U(1)_J$
charge\footnote{This is the $U(1)$ subgroup of $SU(2)_R$ as defined in
\cite{seiw}.}.

\subs{The $\ZZ_2$ case}

In this case the field theory is known, so a detailed comparison can be
made. The Kaluza-Klein harmonics can be obtained by performing the $\ZZ_2$
projection as described above. For example, using the results in table 1 we 
have in the ${\cal N}=4$ theory modes in the ${\bf 20^\prime}$
representation of 
$SU(4)$. This representation is decomposed as 
\eq
{\bf 20^\prime} = \rep{3}{3}{0} \oplus \rep{2}{2}{2} \oplus \rep{2}{2}{-2}
\oplus 
\rep{1}{1}{4} \oplus \rep{1}{1}{-4} \oplus \rep{1}{1}{0} 
\qe
under the global symmetry group $SU(2)\times SU(2)_R\times U(1)_R$. 
The modes get a phase $\exp{(i\pi Q_R/2)}$ under the orientifold action, so  
the invariant representations are those with the $U(1)$ charge $Q_R=0$ mod
$4$. Thus one finds that only $4$ representations are left:
\eq
{\bf 20^\prime} \supset \rep{3}{3}{0} \oplus \rep{1}{1}{4} \oplus
\rep{1}{1}{-4} 
\oplus \rep{1}{1}{0} .
\qe

In case of the operators arising from the second rank gauge fields one has
to take care to account for the ``extra'' phase coming from the orientifold
action on these fields, as discussed earlier. 
These operators 
come from the representations ${\bf 10}$ and ${\bf 45}$ of $SU(4)$, which
decompose as
\eqr
\label{deco}
{\bf 10} &=& \rep{1}{3}{2} \oplus \rep{3}{1}{-2} \oplus
\rep{2}{2}{0}\nonumber \\
{\bf 45} &=& \rep{4}{2}{-2} \oplus \rep{2}{4}{2} \oplus 
\rep{2}{2}{2} \oplus \rep{2}{2}{-2} ,\nonumber \\
&\oplus& \rep{3}{1}{-4} \oplus \rep{1}{3}{4} \oplus 
\rep{3}{3}{0} \oplus \rep{3}{1}{0} \oplus \rep{1}{3}{0} .
\rqe
The orientifold action on the two-form fields is in this case
\eq
\ve B \\ \tilde{B} \ev = - \ve B \\ \tilde{B} \ev 
\qe
where $B$ is the NS-NS two-form field, and $\tilde{B}$ is the R-R
two-form. 
Thus each representation in the decomposition (\ref{deco}) gets in total a  
phase $\exp{(i\pi (Q_R + 2)/2)}$. Thus in this case only operators with
$Q_R=2$ mod $4$ remain. Specifically, in the decomposition of ${\bf 10}$
only the first two 
representations are invariant and the last one drops out, whereas in the
case of the ${\bf 45}$ the first four survive and the remaining ones are
projected out. 

Proceeding this way one finds the results summarized in table 2a below:

\begin{center}
\vspace{.2cm}
\begin{tabular}{|c|c|c|}
\hline
$\Delta$ & $SU(4)$ & $SU(2)\times SU(2)_R\times U(1)_R$
\\ \hline\hline 
2 & $ {\bf 20'} $  & $\rep{3}{3}{0} \oplus \rep{1}{1}{4} \oplus \rep{1}{1}{-4} \oplus \rep{1}{1}{0}$\\
\hline
3 & $ {\bf 10} $  & $\rep{3}{1}{-2}\oplus\rep{1}{3}{2}$   \\ 
\hline
3 & $ {\bf 50} $  & $\rep{2}{2}{4}\oplus\rep{2}{2}{-4}\oplus\rep{4}{4}{0}
\oplus \rep{2}{2}{0}$\\ 
\hline
4 & ${\bf 1}$ & $\rep{1}{1}{0}$ \\
\hline
4 & $ {\bf 45} $  & $\rep{4}{2}{-2} \oplus \rep{2}{4}{2} \oplus 
\rep{2}{2}{2} \oplus \rep{2}{2}{-2}$\\
\hline 
4 & $ {\bf 105} $  &
$\rep{1}{1}{0} \oplus \rep{1}{1}{4} \oplus \rep{1}{1}{-4} \oplus
\rep{1}{1}{8} \oplus \rep{1}{1}{-8}$ \\ 
& &$\oplus \rep{3}{3}{0} \oplus
\rep{3}{3}{4} \oplus \rep{3}{3}{-4} \oplus \rep{5}{5}{0}$\\
\hline  
\end{tabular}

\vspace{.2cm}
{\bf Table 2a: The $\ZZ_2$ case: invariant operators.}
\end{center}
\vspace{.1cm}

The $\rep{1}{1}{0}$ operator here is the dilaton. 

For dimension $\Delta=2$ only the $\rep{3}{3}{0}$ and
$\rep{1}{1}{4}$ modes have charges satisfying (\ref{charge}), so only they 
correspond to chiral primary fields. In fact one can easily identify these
operators in the field theory. In the present case one can use the
perturbative field theory description of the worldvolume dynamics 
\cite{sen,douls}. The low energy degrees of freedom of this field theory
include (in ${\cal N}=1$ language) the vector multiplet $W_\alpha$ and the
chiral multiplets $\phi$ in the adjoint representation of $Sp(2N)$ and 
$V$, $\tilde{V}$ in the antisymmetric representation. These fields
correspond 
to the geometric distances $z$, $v$, $\tilde{v}$ introduced in section 3. 
Under the global
symmetry group $SU(2)\times SU(2)_R\times U(1)_R$ these transform as 
\eqr
\phi \sim \rep{1}{1}{2} \nonumber \\
V \sim \rep{2}{2}{0} \nonumber \\
W_\alpha \sim \rep{1}{2}{1}
\rqe
The operators $\rep{3}{3}{0}$ and 
$\rep{1}{1}{4}$ can be represented as $Tr(V^2)$ and $Tr(\phi^2)$. 
For example \footnote{The
normalization here is such that for the fundamental of $SU(2)$ J assumes
values $-1,1$.} $\rep{3}{3}{0}$ has $R=0$ and $J=2$, so (\ref{charge}) is
satisfied. In the case of 
$\rep{1}{1}{-4}$ however one has $R=-4, J=0$, so (\ref{charge}) is not
satisfied. 

Carrying out this argument for dimensions up to $4$ one reaches the
conclusions summarized in Table 2b. This table, as well as the following
ones, contains only chiral primaries; anti-chiral primary operators will
also be present in the spectrum (and protected) and are conjugates of the
ones listed. 

\begin{center}
\vspace{.2cm}
\begin{tabular}{|c|c|c|}
\hline
$\Delta$ & $SU(2)\times SU(2)_R\times U(1)_R$ & Field content \\ 
\hline\hline 
$ 2 $  & $\rep{1}{1}{4}$ & $Tr(\phi^2)$  \\ 
\hline
$ 2 $  & $\rep{3}{3}{0}$ & $Tr(V^2)$  \\ 
\hline
$ 3 $  & $\rep{1}{3}{2}$ & $Tr(W^2)$  \\ 
\hline
$ 3 $  & $\rep{4}{4}{0}$ & $Tr(V^3)$ \\
\hline
$ 3 $  & $\rep{2}{2}{4}$ & $Tr(V\phi^2)$ \\
\hline
$ 4 $  & $\rep{2}{4}{2}$ & $Tr(W_\alpha W^\alpha  V)$ \\
\hline
$ 4 $  & $\rep{1}{1}{8}$ & $Tr(\phi^4)$  \\ 
\hline
$ 4 $  & $\rep{5}{5}{0}$ & $Tr(V^4)$  \\ 
\hline
$ 4 $  & $\rep{3}{3}{4}$ & $Tr(V^2 \phi^2)$  \\ 
\hline
\end{tabular}

\vspace{.2cm}
{\bf Table 2b: The $\ZZ_2$ case: chiral primary fields.}
\end{center}
\vspace{.1cm}

As in other cases studied in the literature, it was possible to identify
uniquely 
all the surviving Kaluza-Klein states with appropriate operators on the
gauge theory side.

\subs{The Cases with Exceptional Global Symmetry}

This section presents the results for the cases of $\ZZ_3$, $\ZZ_4$ and
$\ZZ_6$, for which the theory on the D3-brane worldvolume has global
exceptional symmetry $E_6$, $E_7$, $E_8$ respectively.

For the $\ZZ_3$ case the invariant operators are those with the $U(1)$ charge
$Q=0$ mod $6$, except for the operators coming from the ${\bf 10}$ and
${\bf 45}$. In the present case the orientifold action on the two-form
fields is 
\eq
\ve B \\ \tilde{B} \ev = \ar -1 & -1 \\ 1 & 0 \ra \ve B \\ \tilde{B} \ev .
\qe
The surviving modes will be linear combinations of these:
\eqr
B^\prime &=& e^{-2\pi i/3} B + \tilde{B} ,\nonumber \\
\tilde{B}^\prime &=& e^{+2\pi i/3} B + \tilde{B} .
\rqe
These combinations transform multiplicatively. The total phase is then
$\exp{(i\pi (Q_R\pm 2)/3)}$, so the surviving operators in the ${\bf 10}$
and ${\bf 45}$ are those with $Q_R=\pm 2$ mod $6$. This way one finds the 
invariant operators given in Table 3a 
below:   

\begin{center}
\vspace{.2cm}
\begin{tabular}{|c|c|c|}
\hline
$\Delta$ & $SU(4)$ & $SU(2)\times SU(2)_R\times U(1)_R$
\\ \hline\hline 
2 & $ {\bf 20'} $  & $\rep{3}{3}{0} \oplus \rep{1}{1}{0}$\\
\hline 
3 & $ {\bf 10} $  & $\rep{3}{1}{-2}\oplus\rep{1}{3}{2}$   \\ 
\hline
3 & $ {\bf 50} $  & $\rep{1}{1}{6} \oplus \rep{1}{1}{-6} \oplus
\rep{2}{2}{0}\oplus \rep{4}{4}{0}$   \\ 
\hline
4 & $ {\bf 45} $  & $\rep{4}{2}{-2} \oplus \rep{2}{4}{2} \oplus 
\rep{2}{2}{2} \oplus \rep{2}{2}{-2}  \oplus
\rep{3}{1}{-4}\oplus \rep{1}{3}{4}$ \\
\hline 
4 & $ {\bf 105} $  &
$\rep{1}{1}{0} \oplus \rep{2}{2}{6} \oplus \rep{2}{2}{-6} \oplus
\rep{3}{3}{0} \oplus \rep{5}{5}{0}$ \\
\hline  

\end{tabular}

\vspace{.2cm}
{\bf Table 3a: The $\ZZ_3$ case: invariant operators.}
\end{center}
\vspace{.1cm}

As in the $\ZZ_2$ case, to identify the chiral primaries one has to check
whether the relation (\ref{charge}) is satisfied. This leads to the
conclusions summarized in Table 3b. 

\begin{center}
\vspace{.2cm}
\begin{tabular}{|c|c|}
\hline
$\Delta$ & $SU(2)\times SU(2)_R\times U(1)_R$ \\ 
\hline\hline 
$ 2 $  & $\rep{3}{3}{0}$ \\ 
\hline
$ 3 $  & $\rep{1}{1}{6}$ \\ 
\hline
$ 3 $  & $\rep{1}{3}{2}$ \\
\hline
$ 3 $  & $\rep{4}{4}{0}$ \\
\hline
$ 4 $  & $\rep{1}{3}{4}$ \\
\hline
$ 4 $  & $\rep{2}{4}{2}$ \\
\hline
$ 4 $  & $\rep{2}{2}{6}$ \\
\hline
$ 4 $  & $\rep{5}{5}{0}$ \\
\hline
\end{tabular}

\vspace{.2cm}
{\bf Table 3b: The $\ZZ_3$ case: chiral primary fields}
\end{center}
\vspace{.1cm}

The analysis for the remaining cases proceeds analogously. 
In the $\ZZ_4$ case the invariant operators are those with the $U(1)$ charge
$Q=0$ mod $8$, again with the exception of operators coming from the
${\bf 10}$ and 
${\bf 45}$. In this case the orientifold action on the two-form
fields is 
\eq
\ve B \\ \tilde{B} \ev = \ar 0 & -1 \\ 1 & 0 \ra \ve B \\ \tilde{B} \ev .
\qe
The surviving modes will be linear combinations of these:
\eqr
B^\prime &=& B + i \tilde{B} ,\nonumber \\
\tilde{B}^\prime &=& B - i\tilde{B}.
\rqe
The total phase is 
$\exp{(i\pi (Q_R\pm 2)/4)}$, so the surviving operators in the ${\bf 10}$ 
and ${\bf 45}$ are those with $Q_R=\pm 2$ mod $8$. This leads to the
results in tables 4a and 4b.

\begin{center}
\vspace{.2cm}
\begin{tabular}{|c|c|c|}
\hline
$\Delta$ & $SU(4)$ & $SU(2)\times SU(2)_R\times U(1)_R$ \\ 
\hline\hline 
2 & $ {\bf 20'} $  & $\rep{3}{3}{0} \oplus \rep{1}{1}{0}$\\
\hline 
3 & $ {\bf 10} $  & $\rep{3}{1}{-2}\oplus\rep{1}{3}{2}$   \\ 
\hline
3 & $ {\bf 50} $  & $\rep{2}{2}{0} \oplus \rep{4}{4}{0}$   \\ 
\hline
4 & $ {\bf 45} $  & $\rep{4}{2}{-2} \oplus \rep{2}{4}{2} \oplus 
\rep{2}{2}{2} \oplus \rep{2}{2}{-2}$ \\
\hline 
4 & $ {\bf 105} $  &
$\rep{1}{1}{0} \oplus \rep{1}{1}{8} \oplus \rep{1}{1}{-8} \oplus
\rep{3}{3}{0} \oplus \rep{5}{5}{0}$
\\
\hline  

\end{tabular}

\vspace{.2cm}
{\bf Table 4a: The $\ZZ_4$ case: invariant operators.}
\end{center}
\vspace{.1cm}

\begin{center}
\vspace{.2cm}
\begin{tabular}{|c|c|}
\hline
$\Delta$ & $SU(2)\times SU(2)_R\times U(1)_R$ \\ 
\hline\hline 
$ 2 $  & $\rep{3}{3}{0}$ \\ 
\hline
$ 3 $  & $\rep{1}{3}{2}$ \\ 
\hline
$ 3 $  & $\rep{4}{4}{0}$ \\ 
\hline
$ 4 $  & $\rep{2}{4}{2}$ \\
\hline
$ 4 $  & $\rep{1}{1}{8}$ \\
\hline
$ 4 $  & $\rep{5}{5}{0}$ \\
\hline
\end{tabular}

\vspace{.2cm}
{\bf Table 4b: The $\ZZ_4$ case: chiral primary fields}
\end{center}
\vspace{.1cm}

Finally in the $\ZZ_6$ case the invariant operators are those with $U(1)$
charge $Q=0$ mod $12$, again with the exception of operators coming
from the ${\bf 10}$ and 
${\bf 45}$. The orientifold action on the two-form
fields is 
\eq
\ve B \\ \tilde{B} \ev = \ar 0 & -1 \\ 1 & 1 \ra \ve B \\ \tilde{B} \ev .
\qe
The total phase turns out to be 
$\exp{(i\pi (Q_R\pm 2)/6)}$, so the surviving operators in the ${\bf 10}$ 
and ${\bf 45}$ are those with $Q_R=\pm 2$ mod $12$. This leads to the
results in tables 5a and 5b.

\begin{center}
\vspace{.2cm}
\begin{tabular}{|c|c|c|}
\hline
$\Delta$ & $SU(4)$ & $SU(2)\times SU(2)_R\times U(1)_R$
\\ \hline\hline 
2 & ${\bf 20'} $  & $\rep{3}{3}{0} \oplus \rep{1}{1}{0}$\\
\hline 
3 & $ {\bf 10} $  & $\rep{3}{1}{-2}\oplus\rep{1}{3}{2}$   \\ 
\hline
3 & ${\bf 50} $  & $\rep{2}{2}{0}\oplus\rep{4}{4}{0}$   \\ 
\hline
4 & $ {\bf 45} $  & $\rep{4}{2}{-2} \oplus \rep{2}{4}{2} \oplus 
\rep{2}{2}{2} \oplus \rep{2}{2}{-2}$ \\
\hline 
4 & ${\bf 105}$  & $\rep{1}{1}{0} \oplus \rep{3}{3}{0}\oplus \rep{5}{5}{0}$\\
\hline  
\end{tabular}
\vspace{.2cm}

{\bf Table 5a: The $\ZZ_6$ case: invariant operators.}
\end{center}
\vspace{.1cm}

\begin{center}
\vspace{.2cm}
\begin{tabular}{|c|c|}
\hline
$\Delta$ & $SU(2)\times SU(2)_R\times U(1)_R$ \\ 
\hline\hline 
$ 2 $  & $\rep{3}{3}{0}$ \\ 
\hline
$ 3 $  & $\rep{1}{3}{2}$ \\
\hline
$ 3 $  & $\rep{4}{4}{0}$ \\
\hline
$ 4 $  & $\rep{2}{4}{2}$ \\
\hline
$ 4 $  & $\rep{5}{5}{0}$ \\
\hline
\end{tabular}

\vspace{.2cm}
{\bf Table 5b: The $\ZZ_6$ case: chiral primary fields.}
\end{center}
\vspace{.1cm}

This is a prediction for the marginal and relevant perturbations, based on
the the correspondence advocated in \cite{mal}. Since there is no known
perturbative description of these fixed points it is not possible to give a
representation of these operators here. One feature which is easy to check
is the presence of the operator which corresponds to the parameter on the
Coulomb branch\footnote{We would like to thank Shamit Kachru for a
discussion on this point.} (which appears in the Seiberg-Witten curves of
these 
theories\cite{minn1,minn2}).  In the $Sp(2N)$ case it is $tr(\phi^2)$, so
it is the 
$\Delta=2$, $\rep{1}{1}{4}$ entry in Table 2b. In the $E_6$, $E_7$, $E_8$
cases the corresponding operator has dimension $3$, $4$ and $6$
(essentially for symmetry reasons). Indeed, in the $E_6$ case (Table 3b)
one finds the 
dimension $3$ operator $\rep{1}{1}{6}$, and in the $E_7$ case (Table 4b)
one finds the dimension $4$ operator $\rep{1}{1}{8}$. There is no
$\rep{1}{1}{12}$ operator in Table 5b, since only relevant and marginal
operators are listed there.

\sect{Conclusions}

It is clearly of interest to make use of Maldacena's duality in contexts,
where other methods are not available.  This note reported an analysis of
spectra of superconformal field theories with $8$ supersymmetries
describing the worldvolume dynamics of D3-branes in the presence of
D7-branes and an orientifold plane. These models can be viewed as points in
F-theory moduli space characterized by a
constant expectation value of the dilaton and Ramond-Ramond scalar. 
At these points the base of the K3 fibration becomes $T^2/\ZZ_n$
($n=2,3,4,6$). The worldvolume theories on the D3-branes have flavour
symmetry groups $SO(8)$, E$_6$, E$_7$ and E$_8$.
In the $\ZZ_2$
case where the field theory is known one can make a
detailed comparison of the supergravity prediction with what is expected in 
the field theory, and as in other cases studied previously the spectra
match.  It should be noted however that the operators considered here
account only for the open strings stretched between D3-branes.  In addition
to 
these operators there are those which account for strings stretched between
D3- and D7-branes.  They correspond to states appearing on the supergravity
side.  For instance there are gauge bosons coming from the coinciding
7-branes which couple to flavour currents in the CFT.

The $\ZZ_3$, $\ZZ_4$, $\ZZ_6$ cases provide examples of superconformal systems
without known field theory descriptions. Yet it is still possible to apply
Maldacena's duality! The predictions reported here cannot be compared with
anything known at this time, but perhaps these and similar calculations may
provide information which will help to find a description of such theories.

\vspace{0.5cm}

\begin{center}
  {\bf Acknowledgments}
\end{center}

We would like to thank Oren Bergman, Albion Lawrence, 
Nikita Nekrasov and especially Juan Maldacena for helpful discussions. 
We would also like to thank T. Hauer for pointing out some misprints
in a previous version.


\end{document}